\documentclass[12pt]{article}
\usepackage{mathptmx}
\usepackage{sprinconf}
\usepackage{graphicx}
\usepackage{bm}
\def\be{\begin{equation}}
\def\ee{\end{equation}}
\begin{document}

\title{Topological quantization of the harmonic oscillator}

\author{Francisco  Nettel\dag\footnote{fnettel@nucleares.unam.mx}, 
Hernando  Quevedo\dag\footnote{quevedo@nucleares.unam.mx}}
\affil{\dag\ Instituto de Ciencias Nucleares, Universidad Nacional Aut\'onoma de M\'exico, A.P. 70-543, M\'exico D.F. 04510, M\'exico}

\beginabstract
We present a derivation of the energy spectrum of the harmonic oscillator
by using  the alternative approach of topological quantization. 
The spectrum is derived from the topological invariants of a 
particular principal fiber bundle which can be assigned to any 
configuration of classical mechanics, when formulated according
to Maupertuis formalism.

\endabstract

\section{Introduction}  \label{intro}

Although the idea of using topology to find quantum properties of 
classical systems is not new and has been used very actively 
in the context of diverse monopole and instanton configurations 
\cite{frankel}, a strict proof of the existence of the 
underlying fiber bundle structure was provided only 
recently for classical gravitational fields \cite{patquev}.  
In an attempt to apply topological quantization  
to classical systems with only a finite number of degrees
of freedom, we found that it is necessary to reformulate 
classical mechanics in such a way that the theorem proved
in \cite{patquev} can be applied. Fortunately, that formulation
already exists and is known as Maupertuis approach.
To show the applicability of the main 
theorem of topological quantization in this case, one needs
to solve several technical problems related to the structure
of principal fiber bundles with specific non compact fibers.
This task is still under investigation. In this paper 
we will present a simple example in which the idea of
topological quantization is applied to a simple harmonic   
oscillator. We will show that in fact it is possible to 
derive the energy spectrum of the harmonic oscillator
by analyzing the topological invariants of the corresponding
principal fiber bundle.

\section{The approach of topological quantization}
 \label{first}
 
 The method
of topological quantization  
can be applied to any
field configuration whose geometrical structure allows the existence of
a principal fiber bundle. In the case of gravitational systems with an
infinite number of degrees of freedom, a theorem proves the existence
and uniqueness  
of such a bundle \cite{patquev}. Indeed, the theorem states that
any solution of Einstein's equations 
minimally coupled to any gauge matter field can be represented
geometrically as a principal fiber bundle with spacetime as 
the base space. The structure group (isomorphic to the standard fiber)
follows from the invariance of the metric of the base space 
with respect
to Lorentz transformations, in the case of a vacuum solution, 
or with respect to a transformation of the gauge group, in the 
case of a gauge matter field. The topological invariants of the
corresponding principal fiber bundle lead to a discretization
of the parameters entering the metric of the base space. 
In the approach of topological quantization this is the kind of 
discretization we are interested in.   
For the proof of the above theorem
it is very important that the base space is equipped with 
a metric whose invariance determines the standard fiber.
In the search for a similar structure in classical mechanics
we found that Maupertuis formalism provides a natural metric
which can be used to fix the base space.

Consider a classical conservative system with $n$ degrees of freedom 
described by the Lagrangian (summation over repeated
indices)
\be
L = \frac{1}{2}h_{ij} \dot{q}^i\dot{q}^j - V(q) \ .
\ee
Although the evolution of this system can be completely described
within the Lagrangian formalism by varying the action $S=\int L dt$,
we will use 
Maupertuis formalism which is based upon the reduced action 
\be
S_0 = \int ds \ , \quad{\rm with}\quad ds^2 = 2 (E-V) h_{ij} d {q}^i d{q}^j \ ,
\label{s0}
\ee
where $E$ is the total energy.  The equations
following from the variation $\delta S_0=0$, together with the expression 
for the time parameter in terms of the reduced action, completely
describe the evolution of the system \cite{arnold}. Equation (\ref{s0})
defines the natural metric $g_{ij} =  2 (E-V) h_{ij}$ which we use to
specify the base space $B$. In general, we can see that to any physical 
system in classical mechanics with $n$ degrees of freedom  corresponds 
an $n-$dimensional Riemannian
space $B$ with metric $g_{ij}$. The potential $V(q)$ is used to characterize
different physical systems. Most systems in classical mechanics are
invariant with respect to Galilean transformations. For the sake of 
simplicity, we limit ourselves here to systems which are invariant
with respect to rotations only. Then to each point of the base space
$B$ we can associate a standard fiber $SO(n)$.  
Furthermore, if we identify the structure group $G$ with the Lie 
group $SO(n)$, we have all the constituents of a $2n$-dimensional 
principal fiber
bundle $P$. According to \cite{patquev}, 
the topological quantization of a classical physical system
follows from the investigation of the topological invariants
of $P$. In the present case, 
the only invariant characteristic class \cite{nashsen} is the Euler class
$e(P)$ which is given in terms of the components of the 
curvature 2-form $R^i_{\ j}$ of $B$ and
whose integration yields an integer, say, $n$ ($2m=n)$
\be
 \int e(P) = \frac{(-1)^m}{2^{2m}\pi^m m!}\int \epsilon_{i_1
   i_2 \dots i_{2m}}\mathbf{R}^{i_1}_{\ i_2}\wedge
\mathbf{R}^{i_3}_{\ i_4}\wedge\cdots \wedge
\mathbf{R}^{i_{2m-1}}_{\ i_{2m}} =  n  \ .
\label{euler} 
\ee
Consider the simple example of a free particle, i.e. $V(q)=0$. 
Then, the metric on the base space $B$ is $g_{ij}= E h_{ij}$
and the corresponding trajectories must be straight lines, 
i.e. the metric $g_{ij}$ must be flat. This implies that
there exists a coordinate system 
in which $h_{ij}=\delta_{ij}$ is the
Euclidean metric. For zero curvature the Euler class vanishes 
and $n=0$. We interpret this result as showing that a free
particle is not quantized from the point of view of topological
quantization. This is in accordance with the results of 
canonical quantization in quantum mechanics. In the general 
case $V(q)\neq 0$ we see that $g_{ij}$ is a conformally flat 
metric, with conformal factor $2(E-V)$, for which clearly 
the curvature is non zero, the Euler class does not vanish
and, according to Eq.({\ref{euler}}), the quantization is not
trivial. In the following section we present an explicit
example of non trivial quantization.
 
\section{The harmonic oscillator}   
\label{hosc}

Consider two harmonic oscillators of the same mass $m$ so that
$h_{ij} = m \delta_{ij}$ and the metric components of the base space $B$ read
\be
g_{ij} = 2m \left[ E -\frac{1}{2}k_1 (q ^1) ^2 - \frac{1}{2}k_2( q ^2) ^2\right]
\delta_{ij}:=e^\phi \delta_{ij} \ .
\ee
This physical system is invariant with respect to transformations of the group 
$SO(2)$ which is taken as the structure group and standard fiber of the principal
bundle $P$. The corresponding Euler class can be expressed as (a coma denotes partial
derivative)
\begin{equation}   
\label{euler1}
e(P) = -\frac{1}{2\pi}R^{1}_{\ 2}  = 
\frac{1}{4\pi}(\phi_{,11} + \phi_{,22})dq^1 \wedge dq^2 \ .
\end{equation}
The calculation is straightforward, but the resulting expression is 
quite cumbersome. To simplify the analysis we consider the special case 
$k_2=0$, and let $k_1=k$ and $q^1=q$. Then
\begin{equation}  \label{euleronedim}
\int e(P) = -\frac{k b}{4\pi}\int
\frac{E + \frac{1}{2}kq^2}{(E - \frac{1}{2}kq^2)^2}dq = n \ , 
\end{equation}
where we choose as $b\pi$ the constant resulting from the integration over $q^2$. 
Integrating over $q$ within the interval $[-q_0, q_0]$, we obtain
\begin{equation}   \label{quantcond}
\frac{bq_0}{q_0^2 - a^2} = n \ , \quad \ a^2 =  \frac{2E}{k} \ , 
\end{equation}
where $a$ represents the
classical turning point. This represents a relationship among the parameters 
describing the harmonic oscillator, i.e., the energy $E$, the constant $k$ and $q_0$ that depends on the former two.
This relationship is unique and gives us information about the discrete nature of the system from the point of view of topological quantization. We call this 
the topological spectrum
of the harmonic oscillator. On the other hand, 
the canonical formalism provides us with a unique canonical spectrum  
for the energy of the system, and we aim for a direct relation between this and 
the topological spectrum. This, however, requires an exact definition of quantum
states in the context of topological quantization \cite{tq}, which is beyond 
the scope of the present work. Nevertheless, a simple way to show the
equivalence is to choose the limit of integration $q_0$ as  
\begin{equation}    \label{limit}
q_0 = \frac{1}{C} - \sqrt{\frac{1}{C^2} + a^2} \ , \quad
 C = \frac{2}{b}\left(\frac{E}{\hbar\omega} - \frac{1}{2}\right)
\ , \quad \omega=\sqrt{\frac{k}{m}}
\end{equation}
which for any positive finite value of $C$ reduces the topological spectrum 
(\ref{quantcond}) to the canonical spectrum
$E = \hbar\omega (n+1/2)$.  
An analysis of the limiting cases shows that 
 the choice (\ref{limit}) is physically meaningful. Indeed,
when $E>>\hbar\omega$ then $\frac{1}{C} \to 0$, and $q_0 \to a$, that is to say
$q_0$ tends to the turning point and we recover the classical limit. Moreover, 
in the limit $E \to 0$, the turning point goes to zero and we have $q_0 \to 0$, 
as expected. 

The above results show that it is possible to obtain the canonical 
energy spectrum of the harmonic oscillator by using the approach of 
topological quantization.

\section*{Acknowledgments}
This work was supported by Conacyt, Mexico, grant 48601.

\end{document}